\documentclass[12pt,a4paper]{article}
\usepackage{type1cm}
\usepackage{braket}
\newcommand{\Tr}{\mathrm{Tr}}
\usepackage{amsmath}	
\usepackage[dvips]{graphicx}	
\title{Tachyon potential in $K B c$ subalgebra}
\author{Syoji Zeze\\
Yokote Seiryo Gakuin High School\\
ztaro21@gmail.com}
\date{Apr. 23, 2010}

\begin{document}
\maketitle
\begin{abstract}
We evaluate the classical action and the effective 
tachyon potential of open string field theory
within $K B c$ subalgebra, which is extensively used in
analytic solution for tachyon condensation recently
found by Erler and Schnabl. 
It is found that the level expansion of the string field
terminates at level 3.  We find that the closed string 
vacuum is a saddle point of the classical action.
We also evaluate the effective potential for tachyon field.
The closed string vaccum becomes stable 
by integrationg out an auxiliary  field.
It is found that the effective potential is bounded below 
hence has no runaway direction.  We also argue validity of
simple identity based solution.  
\end{abstract}

\section{Introduction}
After Schnabl's discovery of the analytic solution \cite{Schnabl:2005gv},
open string field theory have established as a non-perturbative
formulation of string theory.  The solution successfully proves
Sen's conjecture \cite{Sen:1998sm} for unstable D-branes.  
The value of the classical action exactly matches with
the D-brane tension.  Absence of open strings around closed string vacuum
is also shown by the so-called homotopy operator \cite{Ellwood:2006ba}. 
The paper \cite{Schnabl:2005gv} also triggers broad 
interest in string field theory \footnote{
A list of recent works is available in a review article \cite{Fuchs:2008cc}
.}.    

Although basic features of Sen's conjecture have already shown,
analytic solution for tachyon condensation is still 
important to investigate various problems 
in string theory.  The solution defines
string field theory around closed string vacuum as 
\begin{equation}
 S'[\Psi] = S[\Psi + \Psi_{Sc}]\label{063334_14Apr10}
\end{equation}
where $\Psi_{Sc}$ denotes Schnabl's solution and 
$S$ is the action of Witten's cubic SFT \cite{Witten:1985cc}. 
In principle, many important problems could be investigated by this action.
For example, one can show the absence of open string, as have been done
in \cite{Ellwood:2006ba}.  Other D-branes such as 
 lower dimensional D-branes or multiple
D-branes are also expected to be constructed as classical solutions.  
More nontrivial issue is closed strings.  Although the action
(\ref{063334_14Apr10}) lacks physical excitation of 
open strings, it is expected that the action can describe closed 
string physics through open/closed duality on the world sheet.   
However, at our knowledge, such 
applications of Schnabl's solution are quite few at present; 
the analysis of the homotopy operator ,  
evaluation of the gauge invariant closed string operator 
\cite{Ellwood:2008jh, Kawano:2008ry,Kishimoto:2008zj} 
and the boundary state \cite{Kawano:2008jv, Kiermaier:2008qu}. 
One reason which prevent us from extensive research around
closed string vacuum is the complexity of  Schnabl's solution.
It is given by 
\begin{equation}
\Psi_{Sc} = \lim_{N\rightarrow\infty} \left[
\psi_{N} - \sum_{n=0}^{N} \partial_{n} \psi_{n} 
\right],
\end{equation}
where $\psi_n$ is a string field composed by particular insertions
of conformal ghost and anti ghost.  The existence of of isolated piece
$\psi_n$, called phantom piece, forces us to take delicate limit of
large $N$ in the evaluation of physical quantities such as 
classical action.

Recently, a very simplified version of the classical solution for
tachyon condensation is given by Erler and Schnabl
\cite{Erler:2009uj}.  Their solution is 
given by 
\begin{equation}
 \Psi_{ES} = \int_{0}^{\infty} dt\,  (c + c K B c)
e^{-t( K+ 1)},\label{151830_12Apr10}
\end{equation}
where $K$, $B$ and $c$ are string fields which belong to 
a subspace of the star algebra \cite{Okawa:2006vm, Erler:2006hw,
Erler:2006ww}  which is  called `$K B c$  subalgebra'.  The phantom
term is no more absent in their solution.  The sum 
in Schnabl solution is replaced with an integral over the width of semi-infinite strip. 
Such simplification makes calculations easy therefore 
might be useful to investigate various problem
in SFT.

In this paper, we apply the basis of string field used in \cite{Erler:2009uj}
 to evaluation of the classical and effective tachyon potential. We
 employ same gauge condition as that of Erler-Schnabl solution (\ref{151830_12Apr10}) to string field.  Within the $K B c$ subalgebra, this leaves unique choice
\begin{equation}
 \Psi = \int_{0}^{\infty} dt\,  c f(K) B c
e^{-t( K+ 1)},
\end{equation}
where $f(K)$ is a polynomial of $K$.  The power expansion of $f(K)$
corresponds to level truncation with respect to the `dressed'
$\mathcal{L}_0$ operator which is defined by an anticommutator
of $Q_B$ with the gauge condition.   Then, following similar procedure
shown in \cite{Erler:2009uj}, it is straightforward to
evaluate the classical action for $\Psi$.   Surprisingly, it turns
out that the level truncation stops at finite level if we only allow
a field configuration such that leaves the value of the action finite.

This paper is organized as follows.  In Sec.~2, we give a brief review
of the $K B c$ subalgebra and the basis of string field in which we
 are working.   
Sec.~3 is devoted to evaluation of the 
tachyon potential. In Sec.~4, we give an example of identity based 
solution. Our results are summarized and discussions are given 
in Sec.~5.  

\section{String field in the $K B c$ subalgebra}

\subsection{The $K B c $ subalgebra}
The $K B c $ subalgebra \cite{Okawa:2006vm, Erler:2006hw, Erler:2006ww} is spanned by the string fields
$K$, $B$ and $c$ which satisfy
\begin{equation}
 \{c, B\} =1,\qquad [c, K] = c K c, \qquad [K, B] =0,\label{144136_25Mar10}
\end{equation}
where the (anti)commutator is taken with respect to star product.
We omit the $*$ symbol for star
product as as in \cite{Okawa:2006vm, Erler:2006hw, Erler:2006ww,
Erler:2009uj}. 
In addition, the BRST operator acts on these string field as
\begin{equation}
 Q_{B} c = c K c, \qquad Q_{B} B = K, \qquad Q_{B} K =0.\label{144158_25Mar10}
\end{equation}
With the help of this algebra, the authors of Ref.\ \cite{Erler:2009uj}
found a solution of  equation of motion.  The solution has a very
simple expression as
\begin{equation}
 \Psi = c(1+K) B c \frac{1}{1+K}.\label{144214_25Mar10}
\end{equation}
Using (\ref{144136_25Mar10}) and (\ref{144158_25Mar10}), it is straightforward
to show that this solution satisfies the equation of motion $Q_{B} \Psi +
\Psi^2  =0$.  The authors of \cite{Erler:2009uj} also have shown that
the classical action correctly reproduce D-brane tension.  
In order to do this, they rewrote the $1/(1+K)$ factor in
(\ref{144214_25Mar10}) as
\begin{equation}
 \frac{1}{1+K} = \int_{0}^{\infty} dt\, 
e^{-t} e^{-tK} = \int_{0}^{\infty} dt\, 
e^{-t} \Omega^{t},\label{160959_25Mar10} 
\end{equation}
where $\Omega^{t}$ is the wedge state, 
which represents a semi-infinite strip of width $\pi t/2$
in the sliver frame.  

\subsection{String Field in dressed $\mathcal{B}_0$ gauge}

The authors of Ref.\ \cite{Erler:2009uj} also argued the gauge condition which
their solution obeys.   It was found that their solution is in the
`Dressed $\mathcal{B}_{0}$ gauge': 
\begin{equation}
 \frac{1}{2} \mathcal{B}_{0}^{-}\left[    \Psi (1+K)       \right]
\frac{1}{1+K} =0,\label{151814_25Mar10}
\end{equation}
where $\mathcal{B}_{0}^{-} =  \mathcal{B}_{0} -\mathcal{B}_{0}^{\dagger}
$ is a derivation of the star product.  It is not difficult to check 
 (\ref{144214_25Mar10}) with the help of formulas
\begin{equation}
 \frac{1}{2} \mathcal{B}_{0}^{-} c =0, \quad \frac{1}{2}
  \mathcal{B}_{0}^{-} B =0, \quad
 \mathcal{B}_{0}^{-} K = B.
\end{equation}
We would like to consider general form of string field
in this gauge. It is soon realized that 
a ghost number 1 field in this gauge can be obtained
by a slight modification of Erler-Schnabl solution, i.e.,
\begin{equation}
 \Psi =c f(K) B c \frac{1}{1+K},\label{153249_25Mar10}
\end{equation}
where $f(K)$ is an arbitrary function of $K$\footnote{
Note that this string field does not satisfy even when $f(K)$
is real valued function of $K$.  However, there always exists
`real form' of this string field given by \cite{Erler:2009uj}
\[
 \Psi = \frac{1}{\sqrt{1+K}}c f(K) B c \frac{1}{\sqrt{1+K}}.
\]
As explained in \cite{Erler:2009uj}, this string field is gauge
equivalent to the non-real form.   We use
non-real form for convenience.  
}.  
It is 
not difficult to see this is the unique choice if we
restrict ourself within $K B c$ subalgebra.
For example,   multiplying  (\ref{153249_25Mar10})
by a factor such as  $c B g(K)$ or $B g(K) c$ still keeps
the gauge condition.   However, the string field 
can be reduced to the the original form of (\ref{153249_25Mar10}) again 
by contractions of $B$ and $c$.  
In this paper, we consider a case in which $f(K)$ is given by a polynomial
such as
\begin{equation}
f(K) = \sum_{n=0}^{N} t_{n} K^n.\label{154655_25Mar10} 
\end{equation}
Erler-Schnabl solution (\ref{144214_25Mar10}) corresponds to a choice  
 $t_{0} =
t_{1} =1$ and $t_n =0$ for $n \geq 2$.

Before performing an expansion with respect to $K$, let us solve
the equation of motion with ansatz 
 (\ref{153249_25Mar10}). 
Each term in the equation motion can be brought into
 the form  $c \cdots c \cdots
B c (1+K)^{-1}$ by contractions with respect to $B$ and $c$.
After little calculation we have
\begin{equation}
 Q_{B} \Psi = 
\left[
c K c f(K) -c f(K) c K
\right] B c \frac{1}{1+K},
\end{equation}
\begin{equation}
 \Psi^2 =
\left[
c \frac{f(K)}{1+K} c f(K)
-c f(K) c \frac{f(K)}{1+K}
\right] B c \frac{1}{1+K}.
\end{equation}
Then the equation of motion is 
\begin{equation}
 Q_{B} \Psi + \Psi^2  =
\left[
c \left(K+\frac{f(K)}{1+K}\right)
 c f(K) - c f(K)
c \left(K+\frac{f(K)}{1+K}\right)
\right]=0.\label{145011_15Apr10}
\end{equation}
We find three solutions of (\ref{145011_15Apr10}). 
First one is $f(K)=0$, which represents
perturbative vacuum.  Second solution can be
obtained by canceling two terms in (\ref{145011_15Apr10}) each other.
A solution in this case is given by
\begin{equation}
 f(K) = 1+K,
\end{equation}
which is nothing but Erler-Schnabl solution.  The third one is
obtained by setting $K+f(K)/(1+K)=0$.
\begin{equation}
 f(K) = -K(1+K).\label{223303_20Apr10}
\end{equation}
We give a discussion about this new solution in the end 
of section 3.

\section{The classical and effective potential}

In this section, we evaluate the classical action (or equivalently 
classical potential )
\begin{equation}
 V = \frac{1}{2} \Tr [\Psi Q_{B} \Psi] +\frac{1}{3} \Tr [\Psi^3]
\end{equation}
in our setting of
 string field given by (\ref{153249_25Mar10}). 
  All of the calculation can
be preformed by employing a procedure developed in Ref.\
\cite{Erler:2009uj}.  We evaluate the tachyon potential order by
order with respect to the expansion (\ref{154655_25Mar10}).  We
assign  `level' $n$ to the $n$th order term in $f(K)$, 
since the corresponding string field is an eigenstate of `dressed' $\mathcal{L}_{0}$
operator with eigenvalue $n-1$.\footnote{
The `dressed' $\mathcal{L}_{0}$ is given by 
$
\mathcal{L}\Psi =
 \frac{1}{2} \mathcal{L}_{0}^{-}\left\{    \Psi (1+K)       \right\},
$
where $\mathcal{L}_{0}^{-} = \mathcal{L}_{0} -\mathcal{L}_{0}^{\dagger}$.
}
Let us write the potential up to level $N$ as
\begin{equation}
 V = \frac{1}{2} \sum_{m,m} t_{m} K_{m n } t_{n}
+ \frac{1}{3} \sum_{m,m,p} V_{m n p} t_{m} t_{n} t_{p},
\end{equation}
where the sum over $m, n, p$ is taken from $0$ to $N$.  The coefficients
$K_{m n}$ and $V_{m n p}$ can be obtained by plugging 
 the level $n$ field
\begin{align}
 \psi_n & =  c K^n B c \frac{1}{1+K}\notag \\
& = \lim_{s\rightarrow 0} (-\partial_s)^{n} 
\int_{0}^{\infty} e^{-t} dt\,
  c  \Omega^{s}  B c \Omega^{t}.
\end{align}
into the classical action.  It is easily found that both $K_{m n}$ and $V_{m
n p}$ can be reduced to the the well-known
trace \cite{Okawa:2006vm, Erler:2006hw}
\begin{multline}
g(r_{1}, r_{2}, r_{3}, r_{4}) \equiv
  \Tr[B c \Omega^{r_{1}}c
 \Omega^{r_{2}} c \Omega^{r_{3}} c \Omega^{r_{4}}]\\
 = \frac{\left(r_1+r_2+r_3+r_4\right)^{2}}{4 \pi ^3}
\Biggl(
(r_{1}+r_{2}+r_{4})
\sin \left(\frac{2 \pi  r_1}{r_1+r_2+r_3+r_4}\right)\\
+r_{4}\sin \left(\frac{2 \pi  r_2}{r_1+r_2+r_3+r_4}\right)
+r_{2}\sin \left(\frac{2 \pi  r_4}{r_1+r_2+r_3+r_4}\right)\\
-(r_{1}+r_{4})\sin \left(\frac{2 \pi (r_{1}+r_{2}) }{r_1+r_2+r_3+r_4}\right)
   -(r_{1}+r_{2})\sin \left(\frac{2 \pi (r_{1}+r_{4}) }{r_1+r_2+r_3+r_4}\right)\\
   +r_{1}\sin \left(\frac{2 \pi (r_{1}+r_{2}+r_{3}) }{r_1+r_2+r_3+r_4}\right)
   \Biggr). \label{gtrace}
\end{multline}
Then, $K_{m n}$ and $ V_{m n p}$ is given by
\begin{equation}
 K_{m n} =  \lim_{s_{1}\rightarrow 0,\\ s_{2}\rightarrow 0}
(-\partial_{s_1})^m(-\partial_{s_2})^n
 \int_{0}^{\infty}dt_1 \int_{0}^{\infty}dt_2 e^{-t_1 -t_2} 
h_{2} (t_1, t_2, s_1, s_2 ),
\end{equation}
\begin{multline}
 V_{m n p} = 
\lim_{s_{1}\rightarrow 0, s_{2}\rightarrow 0, s_{3} \rightarrow 0 }
(-\partial_{s_1})^m (-\partial_{s_2})^n (-\partial_{s_3})^p
 \int_{0}^{\infty}dt_1 \int_{0}^{\infty}dt_2  \int_{0}^{\infty}dt_3 \\
e^{-t_1 -t_2 -t_3}
 h_{3} (t_1, t_2, t_3, s_1, s_2, s_3),
\end{multline}
where
\begin{multline}
 h_{2} (t_1, t_2, s_1, s_2 ) = - \lim_{u\rightarrow 0}  \partial_{u}
\Bigl[
g(t_{2}, s_1 + t_{1}, u, s_2 ) - g(t_{2}, s_1, t_{1} + u, s_{2}) \\+ 
 g(t_{2}, s_1, t_{1}, u + s_{2}) 
- g(t_{1}, s_{2} + u, t_{2}, s_1) \\+ 
 g(u, t_{2}, s_1 + t_{1}, s_{2}) - g(u, t_{2}, s_1, t_{1} + s_{2})
\Bigr],
\end{multline}
\begin{multline}
 h_{3} (t_1, t_2, t_3, s_1, s_2, s_3) = 
 g(t_3, s_1 + t_1, s_2 + t_2, s_3) - g(t_3, s_1 + t_1, s_2, t_2 + s_3) \\- 
 g(t_3, s_1, t_1 + t_2 + s_2, s_3) + g(t_3, s_1, t_1 + s_2, t_2 + s_3).
\end{multline}
In \cite{Erler:2009uj}, the integrals in the kinetic term (which
corresponds to $K_{m n}$ in our paper) is performed with the help of 
the reparametrization
\begin{equation}
t_{1} = u v, \quad t_{2} = u (1-v) 
\end{equation}
where $u=t_1+t_2$ parametrize the total width of the semi-infinite strip
in the sliver frame,
which corresponds to the total width of the world sheet.
 Under this reparametrization, the measure of the
integral is transformed as
\begin{equation}
 \int_{0}^{\infty}dt_1 \int_{0}^{\infty}dt_2
\rightarrow \int_{0}^{\infty}du \int_{0}^{1}dv\, u.
\end{equation}
While the cubic term is not calculated in \cite{Erler:2009uj}, we find
that similar reparametrization also works. Thus the integrals in $V_{m n
p}$ can be preformed with the replacement
\begin{equation}
t_1 = u  v_1,  \quad t_2 = u  v_2, \quad t_3 = u (1-v_1 -v_2).
\end{equation}
where $u=t_1 +t_2 +t_3$ also represents the width of the world sheet.
The integration measure can also be rewritten into
\begin{equation}
 \int_{0}^{\infty}dt_1 \int_{0}^{\infty}dt_2 \int_{0}^{\infty}dt_3
\rightarrow 
\int_{0}^{\infty}du \int_{0}^{1}dv_{1} \int_{0}^{1-v_{1}} dv_{2}
\, u^2.
\end{equation}
After integrating out $v_{1}$, $v_{2}$ and $v_{3}$,
we obtain the potential as an integral with 
respect to the width $u$ as
\begin{equation}
 V = \int_{0}^{\infty} du\,
e^{-u} A (u, t_{n}),\label{142318_26Mar10}
\end{equation}
where $A(u, t_{n} )$ is finite order in $u$, possibly includes negative
powers of $u$. 
\subsection{The Classical Potential}

We are now ready to evaluate the classical potential up to arbitrary
level.  In principle, all the calculation can be done by hand, but 
 software such as \textit{Mathematica} is useful to evaluate the 
derivatives and integrals which appear in  $K_{m, n}$
and $V_{m,n,p}$.   We find that
the order of $u$ in the integrand of 
 (\ref{142318_26Mar10}) decreases as level increases. 
To illustrate this, we introduce a notation in which the $u$ dependence
of each coefficients in the potential is manifest.
\begin{equation}
 K_{m, n} = \int_{0}^{\infty} du\, 
e^{-u} k_{m,n } (u), \qquad
V_{m,n,p} = 
\int_{0}^{\infty} du\, 
e^{-u} v_{m,n,p } (u).\label{154633_27Mar10}
\end{equation}
For example, $k_{m,n}(u)$ is evaluated up to level 4 as
\begin{equation}
k_{m,n} (u) =
 \left(
\begin{array}{lllll}
 -\frac{u^3}{2 \pi ^2} & 0 & \frac{3 u}{\pi ^2} & \frac{2 \left(-3+\pi
   ^2\right)}{\pi ^2} & 0 \\
 0 & 0 & 0 & 0 & 0 \\
 \frac{3 u}{\pi ^2} & 0 & 0 & 0 & 0 \\
 \frac{2 \left(-3+\pi ^2\right)}{\pi ^2} & 0 & 0 & 0 & -\frac{48 \pi
   ^2}{u^4} \\
 0 & 0 & 0 & -\frac{48 \pi ^2}{u^4} & -\frac{1536 \pi ^2}{u^5}
\end{array}
\right).
\end{equation}
First coefficient which includes negative power of $u$ is 
\begin{equation}
 k_{43}(u) = k_{34}(u) =-\frac{48 \pi^2}{u^4}.
\end{equation}
This factor is divergent in the potential since an integral
\begin{equation}
 - \int_{0}^{\infty} du\, e^{-u}  u^{-4} \times (48 \pi^2 ) 
\end{equation}
is proportional to Gamma function $\Gamma(-3) $,
which is known to be divergent.  Similarly,  any terms 
with negative power of $u$ diverge due to
poles of Gamma function at negative integer.   
The cubic coefficient $K_{m,n} $ also diverges
as level increases.   However, as seen in table \ref{152434_27Mar10},
negative power of $u$ is absent up to level 3.\footnote{We first observe divergence at level 6.
} 
Therefore, if we restrict $f(K)$ in (\ref{153249_25Mar10}) to be polynomial in $K$, and also require finite coefficients in the classical action,
its maximal level is three.  
In other words, level truncation terminates at level 3.\footnote{
In \cite{Erler:2009uj}, it is already shown that the 
classical solution terminates at level 1.
} 
This is quite different from the situation in Siegel gauge or Schnabl's 
$\mathcal{B}_0$ gauge, where level expansion doesn't terminate.
\begin{table}[htbp]
 \begin{center}
  \begin{tabular}{|c|c|c|c|}\hline
  $(m, n, p)$& $v_{m, n, p} (u)$ & $(m, n, p)$& $v_{m, n, p} (u)$  \\
   \hline 
  $(0,0,0)$ & $\frac{3 u^5}{4 \pi ^4}$& $(0,0,1)$ & $\frac{\left(-15+\pi
	     ^2\right) u^4}{6 \pi ^4}$ \\\hline
  $(0, 1, 1)$ & $-\frac{\left(-15+\pi ^2\right) u^3}{3 \pi ^4}$
  & $(0,0,2)$ & $-\frac{\left(-15+4 \pi ^2\right) u^3}{3 \pi ^4}$ \\
  \hline
  $(1,1,2)$ &$\frac{2 \left(-15+\pi ^2\right) u}{\pi ^4}$ &
  $(0,2,2)$ & $\frac{2 \left(-15+4 \pi ^2\right) u}{\pi ^4}$\\ \hline
  $(1,2,2)$ & $-\frac{4 \left(-15+\pi ^2\right)}{\pi ^4}$ &
  $(0,0,3)$ & $-\frac{2 \left(-6+\pi ^2\right) u^2}{3 \pi ^2}$ \\ \hline
  $(0,1,3)$ & $-\frac{2 \left(45-9 \pi ^2+\pi ^4\right) u}{3 \pi ^4}$
  & $(1,1,3)$ & $-\frac{4 \left(-45+6 \pi ^2+\pi ^4\right)}{3 \pi ^4}$\\ \hline
  $(0,2,3)$ & $\frac{4 \left(45-15 \pi ^2+2 \pi ^4\right)}{3 \pi ^4}$ & &
	     \\ \hline
  \end{tabular}
 \end{center}
\caption{A complete list of $v_{m, n, p} (u)$ up to level 3. Other
 coefficients are zero. 
}
\label{152434_27Mar10}
\end{table}

The reason for appearance of negative power of $u$ can be understood
by noting the fact that $K$ is replaced with a derivative 
on a CFT correlator.  Typically, it takes form of
\begin{equation}
\lim_{{s\rightarrow 0}}
  \partial_{s} \sin \frac{z}{s+u} \cdots.
\end{equation} 
then yields $u^{{-2}}$ factor.  

According to (\ref{154633_27Mar10}), the final answer for the tachyon
potential is obtained by performing integration with respect to $u$, together
with the $e^{-u}$ factor.  We present the result below for reference.
\begin{multline}
 V =
\frac{1}{\pi^2}
\Biggl( \frac{15 }{64 \pi ^2}t_0^3-\frac{15 t_1 }{16 \pi ^2}t_0^2+\frac{1}{16}
   t_1 t_0^2+\frac{15 }{16 \pi ^2}t_2 t_0^2-\frac{1}{4} t_2
   t_0^2-\frac{1}{12} \pi ^2 t_3 t_0^2 \\+\frac{1}{2} t_3 t_0^2-\frac{3
   }{32} t_0^2+\frac{15 }{16 \pi ^2}t_1^2 t_0-\frac{1}{16} t_1^2
   t_0-\frac{15 }{4 \pi ^2}t_2^2 t_0+ t_2^2 t_0+\frac{3 }{4}t_2
   t_0-\frac{1}{6} \pi ^2 t_1 t_3 t_0 \\-\frac{15 }{2 \pi
   ^2}t_1 t_3 t_0 +\frac{3}{2} t_1 t_3 t_0 +\frac{4}{3} \pi ^2 t_2 t_3 t_0 +\frac{30
   }{\pi ^2}t_2 t_3 t_0 -10 t_2 t_3 t_0+ \pi ^2 t_3 t_0-3 t_3 t_0\\+\frac{15
   }{\pi ^2}t_1 t_2^2 -t_1 t_2^2-\frac{15 }{4 \pi ^2}t_1^2 t_2 +\frac{1}{4}
   t_1^2 t_2-\frac{1}{3} \pi ^2 t_1^2 t_3+\frac{15 }{\pi ^2}t_1^2 t_3 -2
   t_1^2 t_3
\Biggr)\label{214404_27Mar10}
\end{multline}

\subsection{The instability of the classical potential}

The stationary point of the classical potential, given by solutions of
\begin{equation}
\frac{\partial V}{\partial t_{i}} =0 \quad (i=0,1,2,3)
\end{equation}
can be obtained by numerically.   It turns out that there are eight branches.  Among them, we found
four nontrivial, real solutions, which are summarized in 
table ~\ref{tbl:clspot}.  We discard other solutions since they are
trivial (all fields are zero) or give rise to complex value of
the classical action.  
\begin{table}[htbp]
\begin{center}
 \begin{tabular}{ccccc}
 \hline
 $t_0$ & $t_1$ & $t_2$ & $t_3$ & $V$  \\\hline
 0.2703 & $-0.3928$ & $-0.8194$ & $-0.0358$ &$-0.0306$  \\
 0.2175 &   0.1770 & $-0.9237$ & 0.3032 &  0.0078\\
 0.1195 &  0.1593 & $-0.9076$ &    0.3569 & 0.0081\\
 1.0000 &  1.0000& 0.0000 & 0.0000 & $-0.0507$ \\ \hline
 \end{tabular}
\end{center}
\caption{Stationary points of the classical potential}
\label{tbl:clspot}
\end{table}
The closed string vacuum is given by the configuration $(t_0, t_1, t_2,
t_3)=(1,1,0,0)$. The value of $V$ coincides with D-brane tension
$-1/2 \pi^2$ as expected. 
To see whether these stationary points are stable or not, 
we evaluate Hessian matrix at the closed string vacuum
as have been done in  \cite{AldoArroyo:2009hf} for $\mathcal{B}_{0}$
gauge,
\begin{equation}
 H_{i j} = \frac{\partial^2 V}{\partial t_{i}\partial t_{j}}.
\end{equation}
In particular, the closed string vacuum is 
stable(unstable) if all eigenvalues of $H_{i j}$ is
positive(negative). Otherwise, it is a saddle point.   
 The Hessian matrix for the closed string vacuum is
\begin{equation}
 \left(
\begin{array}{cccc}
 180+6 \pi ^2 & -180+12 \pi ^2 & 180-30 \pi ^2 & -180+48 \pi ^2 \\
 -180+12 \pi ^2 & 180-12 \pi ^2 & -180+12 \pi ^2 & 180-12 \pi ^2-12 \pi ^4 \\
 180-30 \pi ^2 & -180+12 \pi ^2 & 180+24 \pi ^2 & 360-120 \pi ^2+16 \pi ^4 \\
 -180+48 \pi ^2 & 180-12 \pi ^2-12 \pi ^4 & 360-120 \pi ^2+16 \pi ^4 & 0
\end{array}
\right),
\end{equation}
and eigenvalues are found to be 
\begin{equation}
1487.14,\quad -1261.59, \quad 412.919, \quad 79.1831.
\end{equation}
Thus, the Hessian matrix has a negative eigenvalue while others are positive.  This concludes that the closed string vacuum of Erler and Schnabl is a
{\it saddle point} of the classical action, as similar
to the result of \cite{AldoArroyo:2009hf}.  

In addition to the above result, we have found following facts.
\begin{itemize}
 \item We have calculated eigenvalues of Hessian matrix for all 
       solutions in table \ref{tbl:clspot}.  It turns out that
       \textit{all of the four solutions are saddle points}. 
 \item We have checked that each solution satisfies equation of motion
       contracted with itself, i.e. $\Tr[\Psi Q_B \Psi + \Psi^3]=0$
       holds within accuracy of numerical evaluation. 
\end{itemize}

\subsection{Effective Potential}

\begin{figure}[htbp]
\centerline{ \includegraphics{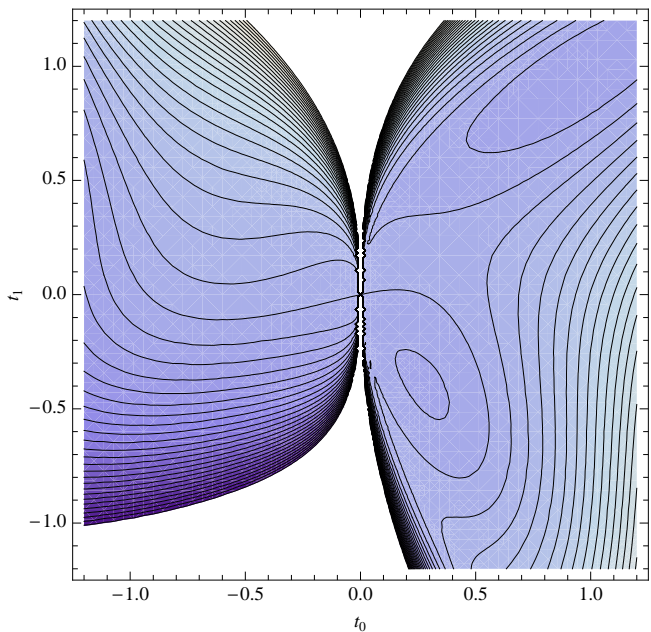}}
\caption{A contour plot of the potential $V_2$ obtained by
integrating out $t_2$ and $t_2$.}
\label{fig:contour}
\end{figure}

The effective potential for tachyon can be obtained by eliminating 
fields other than tachyon field from the classical potential by
solving equations of motion.  
In this paper, we identify $t_0$ as the tachyon mode,
 also such
choice is ad-hoc.   Therefore, in order to obtain
an effective potential as a function of $t_0$,
we have to solve
\begin{equation}
 \frac{\partial V}{\partial t_i} =0 
\qquad(i=0,1,2,3)
\end{equation}
for other fields $t_1, t_2, t_3$. In general, there appear many branches 
since the equations of motion is cubic in each $t_i$.  However, 
one can see that  the classical action (\ref{214404_27Mar10}) is linear in $t_3$.  
Therefore, $t_3$ is an auxiliary field and can be eliminated from the classical 
action by imposing a constraint
\begin{equation}
 \frac{\partial V}{\partial t_3} =0. \label{000618_29Mar10} 
\end{equation}
Furthermore,  we find that the above constraint is again linear 
in $t_2$, because there are no $t_2^2 t_3$ term in the classical action.
Therefore, by solving (\ref{000618_29Mar10}) for $t_2$ and plugging it
to the classical potential,  we obtain an unique
potential which only depends on $t_0$ and $t_1$.
We denote this potential $V_2 (t_0, t_1)$ and
present a contour plot of this potential in figure \ref{fig:contour}.  

We can still find stationary points of $V_2$ 
by numerical method.  Again, we found four real, nontrivial 
stationary points among eight branches. We summarize the obtained
solutions in table \ref{table:v2}, together 
with the value of the potential and their stability 
deduced from the Hessian analysis. 
\begin{table}[htbp]
\begin{center}
 \begin{tabular}{|c|c|c|}\hline
 $(t_0, t_1)$& $V_2$ & stability \\ \hline
 $(0.270333, -0.392821)$ & $-0.0306387$ & stable\\ \hline
 $(0.217487, 0.176975)$ & $0.00784555$ & saddle point\\ \hline
 $(0.119477, 0.159294)$ & $0.00807677$ & unstable\\\hline
$(1.000000,  1.000000)$ & $-0.0506606$ & stable\\ \hline 
\end{tabular}
\end{center}
\caption{A summary of the four 
real stationary points of the potential $V_2$.}
\label{table:v2}
\end{table}
The expected closed string vacuum, stable in this case, is in
the last row of table \ref{table:v2}. 
Curiously,  another stable vacuum shallower than the closed string vacuum
appears in the first row.  We will discuss possible interpretation 
of this vacuum later.

\begin{figure}[htbp]
 \centerline{ \includegraphics{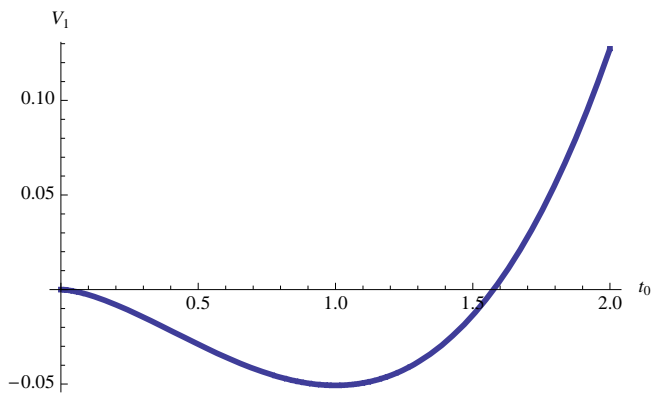}}
 \caption{A plot of the potential $V_1$. We have chosen a branch
 which connects with the closed string vacuum.}
 \label{fig:1dpotential}
\end{figure}

We can further integrate out $t_1$ from $V_2$ by solving
equation of motion.  
In this case, $t_{1}$ cannot be solved uniquely as a function
of $t_{0}$ since  $V_{2}$ is  no more linear in $t_{1}$.
This yields four nontrivial branches.  First, we  would like to 
pick a branch which is 
connected to the closed string vacuum, which is our major concern.  
We can substitute  $t_1$ to the solution of equation of motion
to obtain effective potential $V_1 (t_0)$.  Again the analytic expression
is quite long to show here.   
We show a plot of the potential in Fig. \ref{fig:1dpotential}.

As seen in the plot, the closed vacuum at $t_0 =1$ is stable 
and correctly reproduce the D-brane tension $1/(2 \pi^2) \sim -0.0506 $.
Furthermore, we observed that the potential cannot be extended to negative
value of $t_0$ due to singularity at $t_0 =0$.  
For negative $t_0$, the potential becomes complex valued. 
In earlier  results based on the level truncation in Siegel gauge or $\mathcal{B}_0$
gauge, the potential is valid only in a compact region of tachyon field
, typically starts from slightly before unstable point $t_0 =0$ and 
terminates at some value of $t_0$ \cite{Moeller:2000xv, AldoArroyo:2009hf}
.  Our result seems to be more natural
since potential exactly starts from the unstable vacuum.  
There are no roll off of the tachyon field towards negative value of the tachyon field.  Such feature of the effective
potential completely agrees with the physical picture of unstable D-brane that 
 have been conjecture in \cite{Sen:1998sm}
   ---the closed string vacuum is the
 endpoint of the D-brane decay.

\begin{figure}[htbp]
 \centerline{ \includegraphics[scale=0.7]{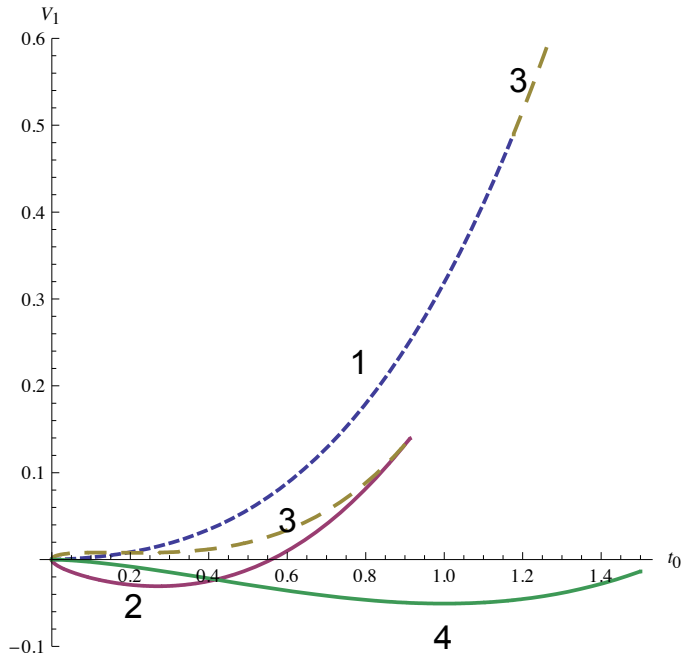}}
 \caption{Four branches of the effective potential $V_1$. 
 We find local minimum in branch 2 and 4.  Branch 3 consists of two
 disconnected curves.  
 }
 \label{fig:branches}
\end{figure}

We can also solve $t_{1}$  in other branches in similar way. 
 A result is shown in fig.~\ref{fig:branches}. 
The branch 4 is the closed string vacuum branch which is already
shown in fig.~\ref{fig:1dpotential}.  It is seen that
none of the branches are extended to the negative $t_0$ region beyond
singularity at origin.  We also found two nontrivial branch points
$t_0 \sim 0.913$ and $t_0 \sim 1.177$ by numerical inspection.
Such branch points are found by scanning a discontinuity
of the imaginary part of $t_1$ as a function of $t_0$. Then we exclude a region in which $t_{1}$ is not real.
One can see that the branch 3 and 2 both terminate
at $t_0 \sim 0.913$, and also another part of branch 3 is
connected with branch 1 at $t_0 \sim 1.177$ smoothly. 
All branches share common features already mentioned
for the closed string vacuum branch; 
neither a runaway direction seen in \cite{Moeller:2000xv},
nor small deviation of tachyon field towards 
negative $t_0$ axis \cite{AldoArroyo:2009hf} is seen.  

In closing this section, let us compare our result from 
level expansion with that of full equation of motion (\ref{145011_15Apr10}).
Both methods share the
string vacuum solution, $(t_0, t_1)=(1,1)$, 
Remaining solutions in level expansion 
have no counterparts in the analysis of full equation of motion. 
It is not surprising since equation of motion in level 
truncated action does not necessary coincides with full
equation of motion.   A rather surprising result is
the alternative solution in full equation motion, $f(K) = -K(1+K)$
in (\ref{223303_20Apr10}).  It is not shown up as a stationary
point of the level expanded action.  Furthermore, it does
not satisfy equation of motion contracted with itself, since
\begin{equation}
\Tr[
\Psi (Q_B\Psi + \Psi^2  )]
= 3 \left(-\frac{15}{\pi ^4}+\frac{1}{\pi ^2}\right).
\end{equation} 
which can be easily calculated from the classical potential
(\ref{214404_27Mar10}).  This indicates that
this solution is ill-defined.  

\section{Identity based solution}
In our setting, there is an identity based solution\footnote{
While completing this paper, 
similar solution, $\Psi = c(1-K)$, appears in 
\cite{Arroyo:2010fq}. 
}
\begin{equation}
 \Psi  = -c K. 
\end{equation}

The equation of motion can be easily checked if one remember
$Q_B c = c K c$.  As is well known, a regularization is needed
to evaluate physical quantity such as classical action.  
We apply a naive regularization via narrow width limit. 
\begin{align}
 \Psi & = -c K \notag \\ 
      & = \lim_{s \rightarrow 0}
       \partial_{s} c \Omega^{s}.  \label{idq}
\end{align}
With this regularization, it is very easy to evaluate the classical action.
First, the quadratic term is given by
\begin{equation}
\Tr \Psi Q_B \Psi = 
- \lim_{s_{1} \rightarrow 0} \lim_{s_{2} \rightarrow 0}
\lim_{u \rightarrow 0}
\partial_{s_{1}} \partial_{s_{2}} 
\partial_{u}
 \Tr[c \Omega^s c \Omega^u c \Omega^s ] .
\end{equation}
The order of limits is important.  The limit with respect to $u$
taken to be first since it originates from $Q_B$, which does not
changes width of the world sheet.  Therefore we take $u\rightarrow 0$
limit first.  Next, we set $s_{1}=s_{2}=s$ before sending them to zero.
It turns out that the trace does not depends on $s$,
so without taking $s\rightarrow 0$  limit we have
\begin{equation}
\Tr \Psi Q_B \Psi = - \frac{1}{2}-\frac{2}{\pi ^2}.\label{idcube}
\end{equation}
Similarly, the cubic term
\begin{equation}
\Tr \Psi^3 = 
 \lim_{s_{1} \rightarrow 0} \lim_{s_{2} \rightarrow 0}
\lim_{s_{3} \rightarrow 0}
 \partial_{s_{1}} \partial_{s_{2}} 
\partial_{s_{3}}
 \Tr[c \Omega^{s_{1}} c \Omega^{s_{2}} c \Omega^{s_{3}}] 
\end{equation}
can be evaluated by setting
$s_{1} = s_{2} = s_{3}$ without taking limits.
\begin{equation}
\Tr \Psi^3 = \frac{2}{9}+\frac{9 \sqrt{3}}{4 \pi ^3}+\frac{\sqrt{3}}{2 \pi }
\end{equation}
The sum of the quadratic term with the cubic term must vanish
since it is an equation of motion contracted with itself.
\begin{equation} 
\Tr \left[
\Psi (Q_{B} \Psi + \Psi^2)
\right].
\end{equation}
However, as is clear from (\ref{idq}) and (\ref{idcube}),
it doesn't vanish.  Therefore $\Psi$ is not a
classical solution in our regularization. 
The reason is easily understood from the fact 
that each traces have different width.  

\section{Discussion}

In this paper, we performed level expansion of string field 
within the $K B c$ subalgebra.    
We find that the level expansion terminates at level 3.  
It is found that the closed string vacuum is a saddle point of
the classical potential.  
As for the effective potential, we confirmed that
it is bounded from below, and exactly starts from perturbative
vacuum.  

An expression of classical action in terms of total strip width $u$,
\begin{equation}
 S = \int_{0}^{\infty} du\,
e^{-u} A(u,t_n),
\end{equation}
is important, since this tells us which width of world sheet 
is most dominant in the classical action.   In principle, 
this kind of expression also appears in
Schnabl solution \cite{Schnabl:2005gv} (given as sums rather integrations)
 and marginal deformation
\cite{Kiermaier:2007ba, Fuchs:2007yy, Kiermaier:2007vu,
Kishimoto:2007bb}, but the multiple integrals or sums with respect to
strip width is very difficult to perform completely.
The simplicity of the $K B c$ subalgebra enable us to perform
multiple integration.   

Tremination of level expansion is also impressive.  Although we
restrict $f(K)$ to be polynomial in this paper,
we can also consider a case of certain series in $K$ such as $f(K) =1/(1+K)$. This
example can be treated by introduction of Schwinger parameter.
It is interesting to evaluate classical action for such string 
field. 

We should note that our result for tahcyon physics is only limited in very 
limited subspace spanned by $K B c$ subalgebra.
Our results may change by inclusion of other modes outside $K B c$ subalgebra.  However, we have done systematic analysis under certain gauge condition, so we believe that our analysis is useful to get insight about physics of tachyon condensation.  Especially, we believe that the effective tachyon potential obtained in this paper 
will help an attempt to derive exact form of the effective
potential, which is not yet available in CSFT.  It will also be
interesting to compare our potential with those derived
from BSFT \cite{Gerasimov:2000zp, Kutasov:2000qp}  
or S-matrix method \cite{Garousi:2007fk,Garousi:2008ge}.

The $K B c$ subalgebra will be very useful for other proposes.
Extension of this subalgebra to fields with nonzero momentum  
will be useful to investigate physics around closed
string vacuum.    Multiple D-branes or lump solutions
will also be interesting.  Application to the gauge invariant
overlap is also important to understand closed sting physics
in terms of open string fields.

\section*{Acknowledgments}

We thank Y.~Okawa and Y.~Hikida for valuable comments at JPS annual
meeting in Okayama University.  We also thank participants of 
`Mini workshop for strings, branes and gauge theory' held at National 
Taiwan University. Especially, we thantk H.~Isono for helpful comments.


\begin{thebibliography}{99}
\bibitem{Schnabl:2005gv}
 M.~Schnabl,
 Adv.\ Theor.\ Math.\ Phys.\  {\bf 10} (2006) 433
 [arXiv:hep-th/0511286].

\bibitem{Sen:1998sm}
 A.~Sen,
 JHEP {\bf 9808} (1998) 012
 [arXiv:hep-th/9805170].

\bibitem{Ellwood:2006ba}
 I.~Ellwood and M.~Schnabl,
 JHEP {\bf 0702} (2007) 096
 [arXiv:hep-th/0606142].

\bibitem{Fuchs:2008cc}
 E.~Fuchs and M.~Kroyter,
 arXiv:0807.4722 [hep-th].

\bibitem{Witten:1985cc}
 E.~Witten,
 Nucl.\ Phys.\  B {\bf 268} (1986) 253.

\bibitem{Ellwood:2008jh}
 I.~Ellwood,
 JHEP {\bf 0808} (2008) 063
 [arXiv:0804.1131 [hep-th]].

\bibitem{Kawano:2008ry}
 T.~Kawano, I.~Kishimoto and T.~Takahashi,
 Nucl.\ Phys.\  B {\bf 803} (2008) 135
 [arXiv:0804.1541 [hep-th]].

\bibitem{Kishimoto:2008zj}
 I.~Kishimoto,
 Prog.\ Theor.\ Phys.\  {\bf 120} (2008) 875
 [arXiv:0808.0355 [hep-th]].

\bibitem{Kawano:2008jv}
 T.~Kawano, I.~Kishimoto and T.~Takahashi,
 Phys.\ Lett.\  B {\bf 669} (2008) 357
 [arXiv:0804.4414 [hep-th]].

\bibitem{Kiermaier:2008qu}
 M.~Kiermaier, Y.~Okawa and B.~Zwiebach,
 arXiv:0810.1737 [hep-th].

\bibitem{Erler:2009uj}
 T.~Erler and M.~Schnabl,
 JHEP {\bf 0910} (2009) 066
 [arXiv:0906.0979 [hep-th]].

\bibitem{Okawa:2006vm}
 Y.~Okawa,
 JHEP {\bf 0604} (2006) 055
 [arXiv:hep-th/0603159].

\bibitem{Erler:2006hw}
 T.~Erler,
 JHEP {\bf 0705} (2007) 083
 [arXiv:hep-th/0611200].

\bibitem{Erler:2006ww}
 T.~Erler,
 JHEP {\bf 0705} (2007) 084
 [arXiv:hep-th/0612050].

\bibitem{AldoArroyo:2009hf}
 E.~Aldo Arroyo,
 JHEP {\bf 0910} (2009) 056
 [arXiv:0907.4939 [hep-th]].

\bibitem{Moeller:2000xv}
 N.~Moeller and W.~Taylor,
 Nucl.\ Phys.\  B {\bf 583} (2000) 105
 [arXiv:hep-th/0002237].

\bibitem{Arroyo:2010fq}
 E.~A.~Arroyo,
 arXiv:1004.3030 [hep-th].
 
 \bibitem{Kiermaier:2007ba}
  M.~Kiermaier, Y.~Okawa, L.~Rastelli and B.~Zwiebach,
  JHEP {\bf 0801} (2008) 028 
  [arXiv:hep-th/0701249].
  
  \bibitem{Fuchs:2007yy}
  E.~Fuchs, M.~Kroyter and R.~Potting,
  JHEP {\bf 0709} (2007) 101 
  [arXiv:0704.2222 [hep-th]].
  \bibitem{Kiermaier:2007vu}
  M.~Kiermaier and Y.~Okawa,
  JHEP {\bf 0911} (2009) 041 
  [arXiv:0707.4472 [hep-th]].
  
\bibitem{Kishimoto:2007bb}
  I.~Kishimoto and Y.~Michishita,
  Prog.\ Theor.\ Phys.\  {\bf 118} (2007) 347 
  [arXiv:0706.0409 [hep-th]].
  
\bibitem{Gerasimov:2000zp}
  A.~A.~Gerasimov and S.~L.~Shatashvili,
  JHEP {\bf 0010} (2000) 034
  [arXiv:hep-th/0009103].
  
\bibitem{Kutasov:2000qp}
  D.~Kutasov, M.~Marino and G.~W.~Moore,
  JHEP {\bf 0010} (2000) 045
  [arXiv:hep-th/0009148].
    
\bibitem{Garousi:2007fk}
  M.~R.~Garousi and E.~Hatefi,
  Nucl.\ Phys.\  B {\bf 800} (2008) 502
  [arXiv:0710.5875 [hep-th]].
  
\bibitem{Garousi:2008ge}
  M.~R.~Garousi and E.~Hatefi,
  JHEP {\bf 0903} (2009) 008
  [arXiv:0812.4216 [hep-th]].
 
\end{thebibliography}
\end{document}